\documentclass{article}
\usepackage[numbers,sort&compress]{natbib}
\usepackage[utf8]{inputenc} 
\usepackage[T1]{fontenc}    

\usepackage{comment}
\usepackage{footnote}
\usepackage{graphicx}
\usepackage{amsmath}
\usepackage{amssymb}
\usepackage{caption}
\usepackage{subcaption}
\usepackage{listings}
\usepackage{amsmath}
\usepackage{hyperref}
\usepackage{multirow}
\usepackage{algorithm}
\usepackage[noend]{algpseudocode}
\usepackage{tikz}
\usetikzlibrary{positioning, arrows.meta, shapes.geometric, decorations.pathreplacing}

\usepackage{xcolor}

\newcommand{\theadb}[1]{\begin{tabular}{@{}c@{}}#1\end{tabular}}

\title{Client-Side Patching against Backdoor Attacks in Federated Learning}
\author{Borja Molina-Coronado\\\texttt{borja.molina@ehu.eus}}

\begin{document}
	
	\maketitle
	
	\begin{abstract}
		Federated learning is a versatile framework for training models in decentralized environments. However, the trust placed in clients makes federated learning vulnerable to backdoor attacks launched by malicious participants. While many defenses have been proposed, they often fail short when facing heterogeneous data distributions among participating clients. In this paper, we propose a novel defense mechanism for federated learning systems designed to mitigate backdoor attacks on the clients-side. Our approach leverages adversarial learning techniques and model patching to neutralize the impact of backdoor attacks. Through extensive experiments on the MNIST and Fashion-MNIST datasets, we demonstrate that our defense effectively reduces backdoor accuracy, outperforming existing state-of-the-art defenses, such as LFighter, FLAME, and RoseAgg, in i.i.d. and non-i.i.d. scenarios, while maintaining competitive or superior accuracy on clean data.
	\end{abstract}

	\section{Introduction}
	
	In recent years, Federated Learning (FL) has emerged as the main paradigm for privacy-preserving machine learning. FL enables collaborative training of a model across multiple clients under the orchestration of a central server \cite{mcmahan2017communication}. The particularity of FL lies in the fact that training is done without sharing the private data stored on the participating clients. This not only ensures data privacy, but reduces data transfers and enables the use of vast amounts of decentralized data to train a model. As a consequence, FL has become the standard learning framework in critical areas such as connected and autonomous vehicles \cite{chellapandi2023federated}, medical diagnosis \cite{antunes2022federated}, and information security \cite{khraisat2024survey}, where strong privacy and security guarantees are required.
		
	However, the decentralized nature of FL makes it susceptible to various threats, including backdoor attacks, which compromise the reliability of the trained models. In backdoor attacks, the attacker is able to manipulate one or more clients of the federation to introduce a hidden functionality into the final model. This hidden functionality is tied to the presence of a \textit{trigger} pattern on the input data, enabling the attacker to control the output of the model. The attack is executed by sending specially crafted model updates to the central server where aggregation is performed \cite{bagdasaryan2020backdoor}.
	
	Backdoor attacks pose a significant threat, particularly in FL applications where the trustworthiness of participating clients cannot be guaranteed. Consequently, practitioners and researchers have focused on developing various defense mechanisms to mitigating these attacks \cite{lyu2022privacy}. Common defense techniques include differential privacy \cite{wei2020federated} and secure aggregation \cite{nguyen2022flame}. However, these approaches often involve trade-offs that can degrade model performance. For example, the majority of existing defense techniques are designed to operate under homogeneous data conditions, assuming that data is independently and identically distributed (i.i.d.) across clients. This assumption, however, does not necessarily hold under realistic settings. As a result, these defenses struggle to maintain acceptable performance, leading to weaker protection against backdoor attacks \cite{molina2024celtibero}. 
	
	Another key characteristic of existing defenses is that the majority are implemented on the server, where only local model updates sent by the clients are accessible. While these updates are useful for identifying irregularities that could indicate malicious activity, placing defenses on the server has several limitations. The most notable lies in the fact that the server cannot access the training data to confirm attacks. Furthermore, since most defenses are applied iteratively during the aggregation process of FL, the aggregated model itself can provide valuable information to attackers, enabling them to craft stealthy attacks that bypass defenses. For instance, attackers can analyze the aggregated model to make malicious updates resemble benign patterns \cite{bagdasaryan2020backdoor,zhang2022neurotoxin}.
	
	To address the limitations of server-side approaches, in this paper we explore a novel client-side defense mechanism. Our defense leverages continuous adversarial learning to identify hidden triggers in the model that could be exploited by attackers, coupled with a patching mechanism aiming at neutralizing such vulnerable spots without compromising model performance. Both mechanisms provide a lightweight and efficient solution that can be seamlessly integrated into existing FL frameworks to increase the robustness of models. To the best of our knowledge, this is the first work to propose a client-side defense based on adversarial learning to counteract backdoor attacks in an FL setting.

	The main contributions of this work are summarized as follows:
	
	\begin{itemize} 
		\item We propose a client-side continuous adversarial learning optimization procedure designed to identify candidate backdoor triggers injected into the global model during FL operation. 
		\item We introduce a client-side patching mechanism to effectively neutralize backdoor attacks in FL models. 
		\item We evaluate our proposed approach against state-of-the-art backdoor attacks, demonstrating its effectiveness in black-box settings. 
	\end{itemize}
	
	The rest of this article is organized as follows: Section \ref{background} provides a overview of the federated learning process and backdoor attacks targeting FL. Section \ref{acab} details the proposed defense mechanism. Section \ref{experimental} describes our experimental setup, including datasets, FL and attack conditions, and evaluates our defense proposal. Section \ref{related} outlines related work on defending against backdoor attacks in FL. Finally, we conclude the paper in Section \ref{conclusions}.

	\section{Background}\label{background}
	
	This section introduces basic knowledge about the Federated Learning framework and backdoor attacks.
	
	\subsection{Federated Learning}
	
	The concept of Federated Learning (FL) was firstly described in \cite{mcmahan2017communication}. It consist of an iterative learning process where each client updates a shared model using its private data and sends only the model updates (i.e., gradients or parameters) to a central server. The server then aggregates the updates to refine the global model, which is subsequently shared back with the clients. Mathematically, the process is defined as follows:

	Let \( \mathcal{D}_k = \{(x^k, y^k)\} \) denote the local dataset of client \( k \), where \( x^k \in \mathbb{R}^d \) represents the samples on client $k$, and \( y^k \) the corresponding labels of samples $x^k$. 
	
	In traditional centralized training, the model is optimized over the combined dataset \( \mathcal{D} = \bigcup_{k=1}^K \mathcal{D}_k \). In FL, however, on each iteration every client \( k \) trains a model locally using its dataset \( \mathcal{D}_k \), and sends the model update \( \Delta \theta_k \) to the central server. The model update is computed as:
	
	\[
	\Delta \theta_k = \theta_k - \theta,
	\]
	
	where \( \theta_k \) is the locally updated model parameters and \( \theta \) refers to the parameters of the global model. 
	
	After each local training performed on the clients, the server aggregates the updates from all \( K \) clients using the Federated Averaging (FedAvg) algorithm to refine the global model. The FedAvg update rule is given by:
	
	\[
	\theta_{t+1} = \theta_t - \frac{\eta}{K} \sum_{k=1}^K \Delta \theta_k,
	\]
	
	where \( \eta \) is the global learning rate, \( t \) is the current training round, and \( \Delta \theta_k \) is the model update from the \( k \)-th client.
	
	\subsection{Backdoor Attacks on FL}
	
	The assumption that all clients are trustworthy makes FL inherently vulnerable to poisoning attacks, especially to backdoor attacks. In such attacks, an adversary introduces a malicious behavior into the model, that is linked to a specific input (\textit{trigger}) pattern. In the absence of the trigger, the model performs as expected. However, when the trigger is present, the model is manipulated to produce an incorrect, predefined output determined by the attacker. The attack process unfolds as follows:
	
	Given the clean local dataset of an attacker-controlled client \( \mathcal{D} = \{(x_i, y_i)\}_{i=1}^N \), where \( x_i \in \mathbb{R}^d \) are the input features, and \( y_i \in \{1, \dots, C\} \) are the corresponding labels, the attacker generates a poisoned dataset \( \mathcal{D}^p = \{(x_i^p, t)\}_{i=1}^M \), where \( x_i^p = x_i + \delta \) is obtained by adding a predefined trigger \( \delta \in \mathbb{R}^d \) to the clean samples \( x_i \), and \( t \) is the target label chosen by the attacker, with \( t \neq y_i \).
	
	The poisoned dataset \( \mathcal{D}^p \) is combined with the clean data \( \mathcal{D} \) for local training. The local loss function for the attacker is given by:
	
	\[
	\mathcal{L}_{\text{poisoned}}(\theta) = \frac{1}{|\mathcal{D}|} \sum_{(x_i, y_i) \in \mathcal{D}} \ell(f_\theta(x_i), y_i) + \frac{1}{|\mathcal{D}^p|} \sum_{(x_i^p, t) \in \mathcal{D}^p} \ell(f_\theta(x_i^p), t),
	\]
	
	where \( f_\theta \) represents the model, \( \ell(\cdot, \cdot) \) is a loss function (e.g., standard cross-entropy for classification tasks) and, \( |\mathcal{D}| \) and \( |\mathcal{D}^p| \) denote the sizes of the clean and poisoned datasets, respectively.
	
	After local training, the attacker sends their malicious updates \( \Delta \theta^p \) to the server. However, since the model aggregation performed in FL can dilute the effect of poisoned updates, attackers often amplify their updates by a factor \( \alpha \gg 1 \), such that:
	
	\[
	\Delta \theta_{\text{amplified}}^p = \alpha \cdot \Delta \theta^p.
	\]
	
	This amplification ensures that the poisoned updates dominate the aggregation process, even in the presence of defenses such as norm clipping or differential privacy \cite{andrew2021differentially}.
	
	\subsubsection{Stealthy Backdoor Strategies}
	
	To avoid detection by defense mechanisms, attackers may craft their updates to resemble those of benign clients. Since benign updates are inaccessible to the attacker, the global model update between rounds \( t-1 \) and \( t \) can serve as a proxy for benign behavior \cite{bagdasaryan2020backdoor}. The loss function of the attacker then combines a stealth term with the poisoning objective:
	
	\[
	\mathcal{L}_{\text{attack}} = \lambda \cdot \mathcal{L}_{\text{poisoned}}(\theta) + (1 - \lambda) \cdot \mathcal{L}_{\text{stealth}}(\theta),
	\]
	
	where \( \mathcal{L}_{\text{poisoned}}(\theta) \) ensures accuracy on both clean and poisoned data, \( \mathcal{L}_{\text{stealth}}(\theta) \) measures the dissimilarity between benign and poisoned gradients (e.g., using L2 \cite{wang2020attack,lyu2023poisoning} or cosine distance \cite{bagdasaryan2020backdoor}), and \( \lambda \in [0, 1] \) is a hyperparameter that controls the trade-off between effectiveness and stealth.
	
	\section{Defense Mechanism}\label{acab}
	
	As we have seen, the malicious behavior embedded into models by backdoor attacks forces the model to associate a specific trigger pattern \( \delta \) with an attacker-chosen target class \( t \). Our defense aims to detect and neutralize these backdoor triggers by iteratively optimizing potential triggers for source-target class pairs and then patching the models of benign clients with the identified triggers to resist backdoor behavior. To do so, our approach incorporates gradient-based optimization techniques inspired by the Projected Gradient Descent (PGD) \cite{goodfellow2014explaining,mkadry2017towards}, along with adversarial training  \cite{qian2022survey} to enhance the robustness of the final models.

	\subsection{Finding Potential Source-Target Class Pairs}
	
	In real-world \textit{black-box scenarios}, the backdoor triggers and source-target class pairs chosen by the attacker are not explicitly known. Consequently, our defense must identify the plausible source-target pairs involved in the attack and the trigger patterns associated with them. To address this, we propose using an iterative optimization approach, similar to PGD, to generate a candidate trigger pattern for each source-target class pair.
	
	Let \( \mathcal{D} \) denote the local dataset of a benign client of the federation, and \( \mathcal{D}_s \) represent the subset of samples belonging to a specific class \( s \). The set of poisoned samples for a source class \( s \) targeting a class \( t \) is defined as:
	
	\[
	\mathcal{D}^p_{s \to t} = \{ (x + \delta_{s \to t}, t) : x \in \mathcal{D}_s \},
	\]
	
	where \( \delta_{s \to t} \) is the trigger pattern that induces the model to misclassify samples from class \( s \) as class \( t \). We parameterize the trigger pattern as:
	
	\[
	\delta_{s \to t} = m \cdot \gamma,
	\]
	
	where \( m \) is a binary mask matrix that specifies the shape and location of the trigger, and \( \gamma \) is a perturbation matrix determining the intensity of the trigger. This parameterization ensures the trigger remains localized and minimally invasive, as suggested by prior work \cite{yao2019latent,nguyen2020input}.
	
	For each candidate source-target pair \( (s, t) \), where \( s \neq t \), we optimize \( \delta_{s \to t} \) to minimize the following combined loss function:
	
	\[
	\mathcal{L}(\delta_{s \to t}) = \mathcal{L}_{\text{bd}} + \mathcal{L}_{\text{clean}} + \mathcal{L}_{\text{sparsity}},
	\]
	
	where:
	
	\[
	\mathcal{L}_{\text{bd}} = \mathbb{E}_{(x^p, t) \sim \mathcal{D}^p_{s \to t}} \ell(f_\theta(x^p), t),
	\]
	
	\[
	\mathcal{L}_{\text{clean}} = \mathbb{E}_{(x, s) \sim \mathcal{D}_s} \ell(f_\theta(x), s),
	\]
	
	\[
	\mathcal{L}_{\text{sparsity}} = \omega \| \delta_{s \to t} \|_1,
	\]
	
	where \( \mathcal{L}_{\text{bd}} \) measures the misclassification performance of the model on poisoned samples from \( \mathcal{D}^p_{s \to t} \), \( \mathcal{L}_{\text{clean}} \) ensures the performance of the model on clean samples from \( \mathcal{D}_s \), and \( \mathcal{L}_{\text{sparsity}} \) includes a regularization term that encourages the trigger pattern \( \delta_{s \to t} \) to be sparse and minimally invasive.
	
	A key challenge in this process is the high computational cost of optimizing \( \delta_{s \to t} \) for all source-target pairs, especially in datasets with a large number of classes. This makes it impractical to perform an exhaustive search for the actual source-target pairs involved in the attack. To mitigate this, we adopt a heuristic approach to prioritize the most likely source-target pairs involved in backdoor activity. For each source class \( s \), the heuristic selects the target class \( t' \) that minimizes a combined loss:
	
	\[
	t' = \arg\min_{t' \in \mathcal{C}, t' \neq s} \Big[ \mathcal{L}_{\text{bd}}(\mathcal{D}^p_{s \to t'}) + \mathcal{L}_{\text{clean}}(\mathcal{D}_s) \Big],
	\]
	
	where \( \mathcal{C} \) is the set of all classes. This selection balances backdoor effectiveness and clean accuracy, reducing the number of pairs to evaluate.
	
	\subsection{Model Patching}
	
	The previous optimization procedure is performed at each iteration of the FL training process. At the end of this process, a patching step is applied to the global model using the triggers identified and optimized by each client. This patching mechanism is designed to neutralize backdoor triggers by disrupting the malicious associations learned by the model that enable the backdoor attack.
	
	Let \( \mathcal{T}_k = \{\delta_{s \to t}\} \) denote the set of optimized trigger patterns identified during training on client \( k \). We use the triggers in \( \mathcal{T}_k \) to construct a patching dataset \( \mathcal{D}^p \) on clients. Specifically, we apply each trigger \( \{\delta_{s \to t}\} \in \mathcal{T}_k \) only to clean samples from its corresponding source class \( s \) in \( \mathcal{D}_s \).
	
	Next, we use the samples in the patching dataset \( \mathcal{D}^p \), along with their original labels \( y \), and the clean dataset \( \mathcal{D} \) to fine-tune the global model \( f_\theta \) by minimizing the following objective function:
	
	\[
	\mathcal{L}_{\text{patch}}(\theta) = \frac{1}{|\mathcal{D}|} \sum_{(x, y) \in \mathcal{D}} \ell(f_\theta(x), y) + \frac{1}{|\mathcal{D}^p|} \sum_{(x^p, y) \in \mathcal{D}^p} \ell(f_\theta(x^p), y),
	\]
	
	where, \( \ell \) represents a loss function (e.g., the cross-entropy loss). By minimizing \( \mathcal{L}_{\text{patch}}(\theta) \), we adjust the parameters of the model \( \theta \) to reduce the sensitivity to backdoor triggers while maintaining the performance on clean samples.
	
	This patching step ensures that the global model \( f_\theta \) becomes robust to backdoor attacks by fixing the vulnerable spots that are part of backdoor/hidden functionalities. Through the application of this process at the end of the FL training, we reduce computational overhead during the standard training process and ensure that patching information is kept private on the benign nodes, impeding adversaries to guide their attacks towards non-patched areas.
	
	\section{Evaluation}\label{experimental}
	
	In this section, we describe the experimental setup used to evaluate our proposal. We then present the results obtained by our defense and compare them with five state-of-the-art defense mechanisms: medianKrum \cite{colosimo2023median}, foolsgold \cite{fung2020limitations}, FLAME \cite{nguyen2022flame}, LFighter \cite{jebreel2024lfighter}, and RoseAgg \cite{yang2024roseagg}.
	
	\subsection{Setup and Methodology}
	
	\begin{table*}[t]
		\centering
		\caption{Datasets and models used in our evaluations.}
		\resizebox{\textwidth}{!}{%
			\begin{tabular}{ c | c | c | c | c | c }
				Dataset & \theadb{\# Training\\Samples} & \theadb{\# Testing\\Samples} & Model & \# Parameters & \theadb{Target\\Class} \\ \hline
				MNIST & 60K & 10K & CNN & $\sim$165K & 0 \\ 
				Fashion-MNIST & 60K & 10K & CNN & $\sim$165K & 0 (T-shirt/top) \\ \hline
			\end{tabular}
		}%
		\label{tab:DLmodels}
	\end{table*}
	
	\paragraph{Datasets} To assess the effectiveness of our defense, we experiment with two widely used benchmark datasets from the FL literature: the MNIST and Fashion-MNIST. The MNIST dataset \cite{lecun1998mnist} contains 70,000 grayscale images of handwritten digits (0-9). The Fashion-MNIST dataset comprises 70,000 grayscale images of fashion items from 10 categories, such as T-shirts, trousers, and shoes. Table~\ref{tab:DLmodels} summarizes the models and configurations used for each dataset.
	
	\paragraph{Attack Setup} We evaluate our proposal against three state-of-the-art backdoor attacks: the Model Replacement Attack (MRA) \cite{bagdasaryan2020backdoor}, the Distributed Backdoor Attack (DBA) \cite{xie2019dba}, and Neurotoxin \cite{zhang2022neurotoxin}. For each attack, we select a random trigger pattern and assign the target class for each dataset as indicated in the rightmost column of Table~\ref{tab:DLmodels}. 
	
	\paragraph{Evaluation Metrics} The performance of our defense mechanism is measured using two standard metrics commonly used in the FL literature: (1) the main task accuracy (MTA), which reflects the performance of the model on the task it is designed for and, (2) the backdoor accuracy (BA), that evaluates the effectiveness of the attack on the model. The MTA is computed as the proportion of correctly classified clean samples out of the total predictions made by the model, whereas the BA value is calculated as the proportion of poisoned samples with the backdoor trigger that are misclassified into the target class selected for the attack.
	
	\paragraph{FL Setup} To simulate the FL framework, each training dataset is divided into $K$ local datasets, each corresponding to a node in the federation. For the i.i.d scenario, local datasets are balanced in size and class distribution. In contrast, non-i.i.d local datasets are generated by sampling the main dataset according to a Dirichlet distribution \cite{lin2016dirichlet} with \( \alpha = 0.5 \). The federation consists of 100 nodes in the i.i.d scenario and 20 nodes in the non-i.i.d scenario, with 40\% of the nodes acting as adversaries and sending poisoned updates to the server. The overall FL process spans 50 global training (aggregation) rounds. Within each FL round, between 60\% and 90\% of clients are selected at random to perform three local training epochs and send their gradients to the server.

	\subsection{Attack Mitigation Results}\label{results}
	
	The results of experiments conducted on the MNIST and Fashion-MNIST datasets for the i.i.d. scenario are shown in Tables~\ref{tab:mnist-iid} and \ref{tab:fmnist-iid}, respectively. As observed, the accuracy of the models on clean data (MTA) remains stable across most scenarios, whereas BA values for MRA and Neurotoxin attacks vary depending on the defense mechanism applied. Specifically, for the MNIST dataset, MRA and Neurotoxin attacks achieved high BA values (above 90\%) against defenses such as \textit{MedianKrum}, \textit{FoolsGold}, \textit{FLAME}, and \textit{RoseAgg}. In contrast, the DBA attack was effectively neutralized by all defenses in this dataset. Remarkably, the most effective defenses were \textit{LFighter}, which completely neutralized the effects of all attacks, with BA values of 0\%, and our proposed defense mechanism, which significantly reduced the BA of MRA and DBA attacks to 0.6\% and 2.2\%, respectively, while completely eliminating the impact of the Neurotoxin attack.
	
	For the Fashion-MNIST dataset, the inability of defenses such as \textit{FoolsGold} and \textit{RoseAgg} to stop the DBA and MRA attacks is evidenced by the high BA values (above 97\%). \textit{FLAME} was vulnerable to the Neurotoxin attack, with a BA of 70\% and a reduced MTA of 74\%. Again, \textit{LFighter} demonstrated robust performance, with stable values of 86\% for MTA and 0\% BA for all attacks. Strong results were also achieved by our defense, with reduced BA values for MRA and DBA attacks of 3.9\% and 0.7\%, respectively, and complete effectiveness in neutralizing the Neurotoxin attack. Overall, while \textit{LFighter} achieved immunity to all attacks, our defense mechanism substantially reduced BA values to an average of 3\%, while maintaining comparable or superior accuracy on clean data with respect to the baseline model.
	
	\begin{table*}[tb]
		\centering
		\caption{MNIST results for the i.i.d scenario}
		\resizebox{\textwidth}{!}{%
			\begin{tabular}{  l | c  c | c  c | c  c | c  c |}
				\multirow{2}{*}{Defenses} & \multicolumn{2}{|c|}{No attack} & \multicolumn{2}{|c|}{MRA \cite{bagdasaryan2020backdoor}} & \multicolumn{2}{|c|}{DBA \cite{xie2019dba}} & \multicolumn{2}{|c|}{Neurotoxin \cite{zhang2022neurotoxin}} \\ \cline{2-9} 
				& MTA & ASR & MTA & ASR & MTA & ASR & MTA & ASR \\ \hline
				\textit{No Defense} & 0.973 & - & 0.969 & 0.986 & 0.969 & 0.987 & 0.955 & 0.102 \\ \hline 
				\textit{MedianKrum \cite{colosimo2023median}} & 0.969 & - & 0.966 & 0 & 0.967 & 0 & 0.615 & 0.915 \\
				\textit{FoolsGold \cite{fung2020limitations}} & 0.973 & - & 0.968 & 0.986 & 0.962 & 0 & 0.962 & 0 \\
				\textit{FLAME \cite{nguyen2022flame}} & 0.969 & - & 0.968 & 0 & 0.969 & 0 & 0.895 & 0.932 \\
				\textit{LFighter \cite{jebreel2024lfighter}} & 0.974 & - & 0.968 & 0 & 0.970 & 0 & 0.969 & 0 \\
				\textit{RoseAgg \cite{yang2024roseagg}} & 0.972 & - & 0.970 & 0.985 & 0.973 & 0 & 0.967 & 0  \\  \hline
				\textit{This proposal} & 0.973 & - & 0.953 & 0.006 & 0.961 & 0.022 & 0.951 & 0 \\
				
			\end{tabular}
		}%
		\label{tab:mnist-iid}
	\end{table*}	
	
	\begin{table*}[tb]
		\centering
		\caption{Fashion-MNIST results for the i.i.d scenario}
		\resizebox{\textwidth}{!}{%
			\begin{tabular}{  l | c  c | c  c | c  c | c  c |}
				\multirow{2}{*}{Defenses} & \multicolumn{2}{|c|}{No attack} & \multicolumn{2}{|c|}{MRA \cite{bagdasaryan2020backdoor}} & \multicolumn{2}{|c|}{DBA \cite{xie2019dba}} & \multicolumn{2}{|c|}{Neurotoxin \cite{zhang2022neurotoxin}} \\ \cline{2-9} 
				& MTA & ASR & MTA & ASR & MTA & ASR & MTA & ASR \\ \hline
				\textit{No Defense} & 0.867 & - & 0.861 & 0.973 & 0.864 & 0.976 & 0.841 & 0.016 \\ \hline 
				\textit{MedianKrum \cite{colosimo2023median}} & 0.863 & - & 0.853 & 0.973 & 0.865 & 0 & 0.535 & 0.902 \\
				\textit{FoolsGold \cite{fung2020limitations}} & 0.867 & - & 0.861 & 0.975 & 0.862 & 0.977 & 0.852 & 0 \\
				\textit{FLAME \cite{nguyen2022flame}} & 0.865 & - & 0.863 & 0 & 0.865 & 0 & 0.746 & 0.701 \\
				\textit{LFighter \cite{jebreel2024lfighter}} & 0.865 & - & 0.865 & 0 & 0.866 & 0 & 0.860 & 0 \\
				\textit{RoseAgg \cite{yang2024roseagg}} & 0.867 & - & 0.861 & 0.973 & 0.864 & 0.976 & 0.841 & 0.016  \\  \hline
				\textit{This proposal} & 0.867 & - & 0.863 & 0.039 & 0.863 & 0.007 & 0.854 & 0 \\
				
			\end{tabular}
		}%
		\label{tab:fmnist-iid}
	\end{table*}
	
	When evaluating defenses with non-i.i.d. data (see Tables~\ref{tab:mnist-niid} and \ref{tab:fmnist-niid}), our experiments reveal key differences in the performance of the defenses. All evaluated defenses, except for our proposal, have shown to be vulnerable to at least two of the three backdoor attacks tested, with an average BA of 95\%. Among existing methods, \textit{RoseAgg} was found to produce models highly susceptible to all considered backdoor attacks. Notably, \textit{LFighter}, the most effective defense under i.i.d. settings, proved particularly vulnerable to MRA and DBA attacks in this non-i.i.d. scenario, with BA values of 98\%. In contrast, our proposed defense has shown to be robust against all tested attacks. Specifically, in the experiments conducted with the MNIST dataset (see Table~\ref{tab:mnist-niid}), our proposal reduced the BA values for the MRA attack to 6\%. Similarly, on the Fashion-MNIST dataset (see Table~\ref{tab:fmnist-niid}), the obtained BA values remained close to zero, with 0.4\% for MRA and 0.2\% for DBA attacks, respectively. Notably, our defense maintained high accuracy on clean data, yielding performance comparable to or better than state-of-the-art methods in scenarios without attacks.

	\begin{table*}[t!]
		\centering
		\caption{MNIST results for the non-i.i.d. scenario}
		\resizebox{\textwidth}{!}{%
			\begin{tabular}{  l | c  c | c  c | c  c | c  c |}
				\multirow{2}{*}{Defenses} & \multicolumn{2}{|c|}{No attack} & \multicolumn{2}{|c|}{MRA \cite{bagdasaryan2020backdoor}} & \multicolumn{2}{|c|}{DBA \cite{xie2019dba}} & \multicolumn{2}{|c|}{Neurotoxin \cite{zhang2022neurotoxin}} \\ \cline{2-9} 
				& MTA & ASR & MTA & ASR & MTA & ASR & MTA & ASR \\ \hline
				\textit{No Defense} & 0.979  & - & 0.978 & 0.991 & 0.978 & 0.992 & 0.976 & 0.836 \\ \hline 
				\textit{MedianKrum \cite{colosimo2023median}} & 0.958 & - & 0.953 & 0.989 & 0.962 & 0 & 0.951 & 0.480 \\
				\textit{FoolsGold \cite{fung2020limitations}} & 0.980 & - & 0.978 & 0.991 & 0.979 & 0.992 & 0.976 & 0 \\
				\textit{FLAME \cite{nguyen2022flame}} & 0.970 & - & 0.969 & 0.991 & 0.976 & 0 & 0.970 & 0.981 \\
				\textit{LFighter \cite{jebreel2024lfighter}} & 0.979 & - & 0.979 & 0.991 & 0.980 & 0.990 & 0.977 & 0 \\
				\textit{RoseAgg \cite{yang2024roseagg}} & 0.962 & - & 0.974 & 0.988 & 0.968 & 0.983 & 0.977 & 0.989 \\ \hline
				\textit{This proposal} & 0.979 & - & 0.954 & 0.064 & 0.964 & 0.042 & 0.958 & 0 	
			\end{tabular}
		}%
		\label{tab:mnist-niid}
	\end{table*}
	
	\begin{table*}[t!]
		\centering
		\caption{Fashion-MNIST results for the non-i.i.d. scenario}
		\resizebox{\textwidth}{!}{%
			\begin{tabular}{  l | c  c | c  c | c  c | c  c |}
				\multirow{2}{*}{Defenses} & \multicolumn{2}{|c|}{No attack} & \multicolumn{2}{|c|}{MRA \cite{bagdasaryan2020backdoor}} & \multicolumn{2}{|c|}{DBA \cite{xie2019dba}} & \multicolumn{2}{|c|}{Neurotoxin \cite{zhang2022neurotoxin}} \\ \cline{2-9} 
				& MTA & ASR & MTA & ASR & MTA & ASR & MTA & ASR \\ \hline
				\textit{No Defense} & 0.872  & - & 0.869 & 0.984 & 0.858 & 0.988 & 0.857 & 0.564 \\ \hline 
				\textit{MedianKrum \cite{colosimo2023median}} & 0.853 & - & 0.817 & 0.400 & 0.841 & 0 & 0.599 & 0.957 \\
				\textit{FoolsGold \cite{fung2020limitations}} & 0.866 & - & 0.865 & 0.985 & 0.862 & 0.985 & 0.863 & 0 \\
				\textit{FLAME \cite{nguyen2022flame}} & 0.861 & - & 0.860 & 0.985 & 0.871 & 0 & 0.792 & 0.936 \\
				\textit{LFighter \cite{jebreel2024lfighter}} & 0.877 & - & 0.852 & 0.984 & 0.862 & 0.988 & 0.856 & 0.685 \\
				\textit{RoseAgg \cite{yang2024roseagg}} & 0.844 & - & 0.865 & 0.988 & 0.844 & 0.983 & 0.836 & 0.715 \\ \hline
				\textit{This proposal} & 0.877 & - & 0.867 & 0.004 & 0.846 & 0.002 & 0.833 & 0.004 	
			\end{tabular}
		}%
		\label{tab:fmnist-niid}
	\end{table*}
	
	\section{Related Work}\label{related}
	
	Among existing defenses aiming to mitigate the impact of backdoor attacks, techniques such as differential privacy \cite{sun2019can,wei2020federated,huang2024vppfl} and secure aggregation \cite{fung2020limitations,blanchard2017machine,nguyen2022flame,jebreel2023fl,siriwardhana2024shield,yang2024roseagg} have been proposed. Differential privacy solutions add noise to model weights to minimize the impact of malicious updates. The main challenge is finding an adequate level of noise that is sufficient to mitigate attacks without degrading model performance \cite{dwork2006calibrating}. In contrast, secure aggregation mechanisms focus on discarding malicious updates during aggregation.
	
	Recent proposals for secure aggregation include \cite{li2021lomar} and \cite{jebreel2024lfighter}. In \cite{li2021lomar}, an estimation of the density function of updates is used to identify malicious updates as those that fall outside the boundaries of their estimated benign neighborhood. This defense is effective in mitigating label-flipping attacks under different non-i.i.d. conditions. In \cite{jebreel2024lfighter}, a clustering algorithm is proposed to group model updates into two categories, with the denser group being considered malicious. However, these techniques have limitations with respect to the types of attacks (e.g., backdoor attacks) or data scenarios they can handle \cite{molina2024celtibero}.
	
	To improve detection accuracy and, consequently, the robustness of secure aggregators, variations of the federated learning scheme have been introduced \cite{cao2020fltrust,andreina2021baffle,qayyum2022making,rieger2024crowdguard}. These modifications mainly focus on providing contextual information, such as associations between data and parameters \cite{qayyum2022making} or benign client feedback \cite{andreina2021baffle,rieger2024crowdguard}, which enriches detection. However, these approaches incur additional communication costs, a sensitive parameter for some FL applications \cite{yu2021toward}.
	
	Alternative post-aggregation techniques, like neuron pruning, where low-activity neurons with benign data are zeroed out, have also been explored \cite{wu2020mitigating}. Since pruning changes the architecture of the model, these methods are prone to impacting the predictive performance of the final model. Additionally, unlearning has been proposed as a mean to remove malicious influences on the model \cite{wu2022federated}. However, the lack of realistic experimental assumptions makes this approach far from meeting actual needs \cite{gupta2021adaptive}.
	
	In terms of the most similar works to ours, one can find \cite{zhang2024sars} and \cite{pan2024one}. In \cite{zhang2024sars}, the authors propose a FL framework where backdoor knowledge is removed through alignment of the global (backdoored) model and the personalized (clean) model trained on each benign node. The work in \cite{pan2024one} removes backdoors upon aggregation by assuming that training data is available on the server, which is not always feasible. Our work differs from \cite{zhang2024sars} and \cite{pan2024one} in that we propose an adversarial learning method at the client nodes, working iteratively within the FL process. The goal of this method is to craft candidate backdoor triggers that patch and remove actual backdoors introduced by the attacker into the global model. A key distinction is that adversarial knowledge, along with local data, is kept private on each client.

	\section{Conclusions}\label{conclusions}
	
	In this work, we proposed a novel patching mechanism for the client-side of federated learning. Specifically, we introduced an iterative optimization process for candidate trigger patterns using adversarial learning techniques, enabling effective patching and neutralization of highly sophisticated backdoor attacks. Our experimental results demonstrate that the proposed method is highly effective against MRA, DBA, and Neurotoxin attacks, achieving backdoor accuracy values below 3\%, while maintaining comparable or superior performance to the baseline on the main task. Compared to existing defenses, our approach performed similarly to the state-of-the-art defense, LFighter, for i.i.d. data, while significantly outperformed all existing methods in non-i.i.d. scenarios. These results demonstrate that our defense is a robust and versatile solution for a variety of backdoor attacks, particularly in contexts where the composition of local datasets cannot be assumed.

	
	\bibliography{references}
	
	\bibliographystyle{unsrtnat}
	
\end{document}